# Electromagnetically induced transparency lasing in distributed resonant feedback system

Yaoyao Liang,[1,*] Xinyue Wang,[1] Hongyi Tian,[1] Yulong Fan,[1] Qi Qin,[2] Feihu Wang[1,†]

*[1]Quantum Science Center of Guangdong-HongKong-Macao Greater Bay Area, Shenzhen, 518000, China*
*[2]Department of System Engineering, City University of Hong Kong, Hong Kong, 999077, China*

We demonstrate a loss-enabled mechanism that can realize single electromagnetically induced transparency (EIT) analogue in periodic waveguide–resonator structures. Unlike previously reported passband-based EIT analogues, which arise from compressed passbands in the intrinsic-loss-free limit, the proposed mechanism exploits resonator intrinsic loss to generate an isolated transparency mode within the original bandgap. The resulting EIT state remains immune to finite-size mode discretization, enabling robust single-frequency lasing through distributed resonant feedback while requiring a threshold modal gain lower than that of an equally long Fabry–Pérot laser. Our findings extend classical EIT analogues from passive spectral phenomena to active laser operation.

Electromagnetically induced transparency (EIT) arises from destructive interference between excitation pathways, opening a narrow transparency window within a broad absorption band and giving rise to steep dispersion that enables slow-light propagation [1–4]. The emergence of EIT analogues has extended this interference phenomenon from purely quantum systems to classical optical resonant platforms [5-10] operating under ambient conditions, enabling a broad range of integrated photonic applications [11-13], including optical storage [14,15], switching [16-18], and sensing [19-21]. Beyond these well-established applications, the unique combination of ultranarrow linewidth, high transmission, and large side-mode suppression ratio renders EIT modes naturally attractive for robust single-frequency lasing operation. Nevertheless, this potential has remained unexplored, largely because most optical EIT systems have been developed to exploit their high-transmission characteristics rather than to serve as laser cavities, and therefore lack the optical feedback required for laser oscillation. Although periodic waveguide–resonator structures [22-24] inherently provide distributed resonant feedback, previous studies have primarily attributed the EIT response to a narrow passband squeezed between polariton–polariton [22] or polariton–Bragg [23, 24] bandgaps. For a finite structure consisting of N periods, however, this passband is quantized into N−1 discrete longitudinal modes [25-29], fundamentally preventing robust single-frequency lasing.

In this Letter, we theoretically demonstrate a loss-enabled mechanism for realizing an EIT analogue in the original bandgap of a periodic waveguide–resonator structure, rather than in the conventionally compressed passband. We show that resonator intrinsic loss is the key ingredient for the isolated EIT mode. The resulting EIT mode persists even in finite-length structures and supports robust single-frequency lasing through distributed resonant feedback. These findings establish a fundamentally new route toward EIT-based lasers.

We use a classic translationally invariant system in which a waveguide side coupled a side resonator in each unit cell, as shown in Fig.1(a). Each resonator is modeled as a single-mode resonator with resonance frequency $\omega_a$, intrinsic loss rate $\gamma_i$, radiation loss $\gamma_e$ and external coupling rate $\kappa$ to the side-coupled waveguide. The resonators are assumed to be identical, symmetric with respect to the two ports, and free of direct inter-resonator coupling, the fields in the system can be written as

$$E(z) = \left[a(z)e^{iKz} + b(z)e^{-iKz}\right]U(x,y), \qquad (1)$$

the forward and backward field amplitudes before coupling to the $N_{th}$ unit cell is denoted as $a_N$ and $b_N$, $K$ corresponds to Bloch wavevector of the entire system, $U(x,y)$ represents transverse field profile. Using the temporal coupled mode theory [30], the transmission matrix for a waveguide side coupled to a resonator with resonance frequency $\omega_a$ can be calculated as

$$\boldsymbol{M_R} = \begin{bmatrix} 1 - \frac{\gamma_e}{-i\Delta\omega_a+\gamma_i} & \frac{-\gamma_e}{-i\Delta\omega_a+\gamma_i} \\ \frac{\gamma_e}{-i\Delta\omega_a+\gamma_i} & 1 + \frac{\gamma_e}{-i\Delta\omega_a+\gamma_i} \end{bmatrix}, \qquad (2)$$

where $\Delta\omega_a = \omega - \omega_a$ represents the frequency detuning from resonator frequency. The transmission matrix for the pure waveguide section of length $L$ can be written as

$$\boldsymbol{M_P} = \begin{bmatrix} e^{i\beta L} & 0 \\ 0 & e^{-i\beta L} \end{bmatrix}, \qquad (3)$$

where $\beta$ is the propagation constant of the unperturbed waveguide mode. Due to the propagation phase accumulated in a unit cell determines the Bragg resonance condition, we can write the $\beta$ through the Taylor series near the Bragg resonance frequency as

$$\beta(\omega) \approx \beta_0(\omega_b) + \frac{d\beta}{d\omega}|_{\omega_b}\Delta\omega_b = m\frac{\pi}{L} + \frac{\Delta\omega_b}{v_g}, \qquad (4)$$

in which $\Delta\omega_b = \omega - \omega_b$ is the frequency detuning from Bragg frequency and hereafter only the first order Bragg series, m=1, is taken into consideration for the sake of simplicity. The transmission matrix through a unit cell

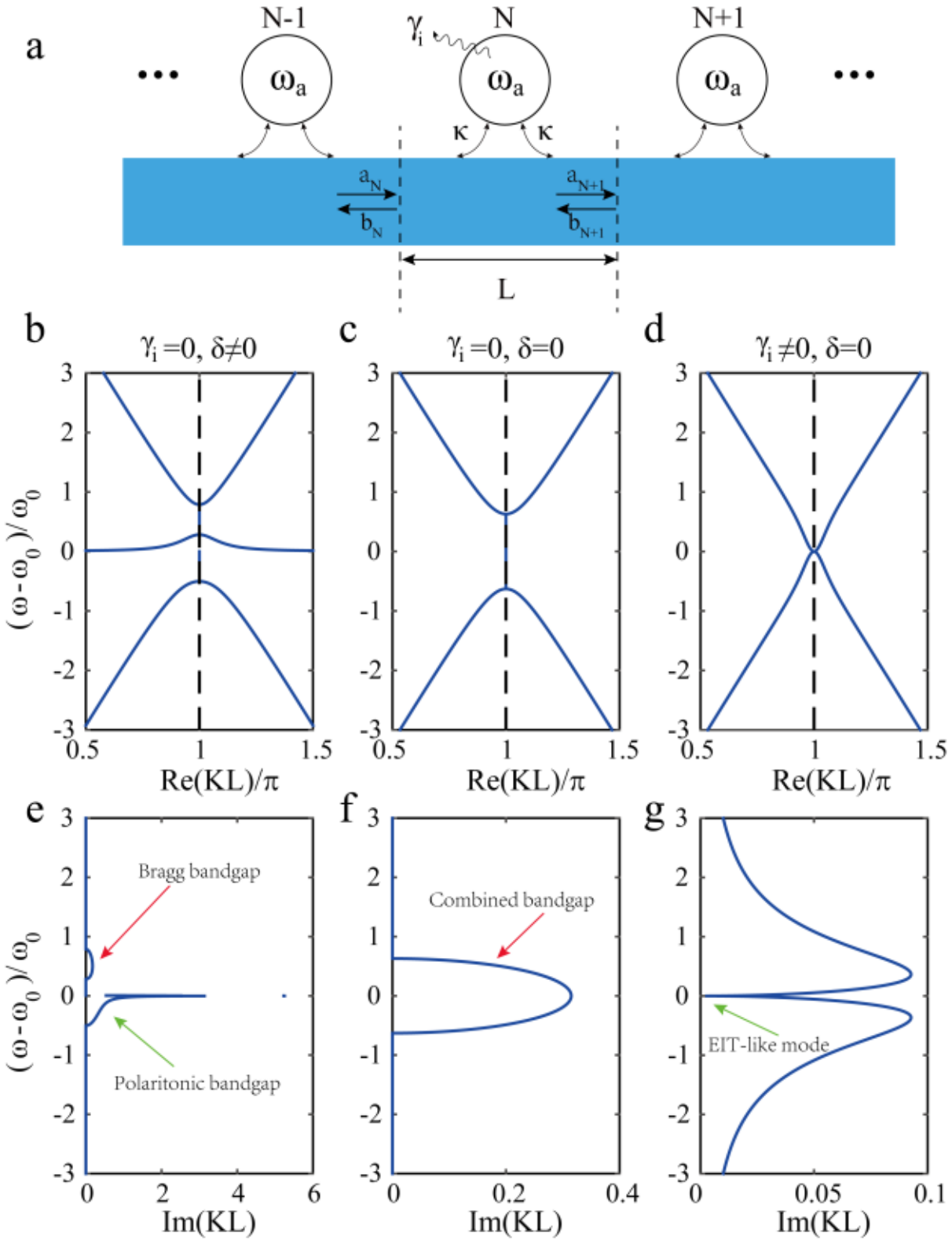

FIG. 1. (a) Sketch of a straight waveguide side-coupled to N resonators with equally spacing of L. The forward and backward optical waves at left side of $N_{th}$ resonator are described by $a_N$ and $b_N$. The resonator has resonant frequency of $\omega_a$, absorption loss $\gamma_i$, radiation loss $\gamma_e$, the coupling strength between resonator and waveguide $\kappa = \sqrt{\gamma_e}$. (b–d) and (e–g) show the real and imaginary parts of $K$L as functions of frequency, respectively. Panels (b, e) and (c, f) correspond to the frequency-misaligned $(\delta = \omega_0(3-\pi))$ and zero-detuning $(\delta = 0)$ cases between the Bragg resonance and the resonator resonance in the absence of intrinsic loss $(\gamma_i = 0)$. Panels (d, g) depict the zero-detuning case in the presence of intrinsic resonator loss ($\gamma_i = 10\gamma_e$). The parameters associated here are: $\gamma_e = 0.1\omega_0, L\omega_0/v_g = 0.5$.

can thus be determined as

$$\boldsymbol{M_U} = \boldsymbol{M_{P/2}} \boldsymbol{M_R} \boldsymbol{M_{P/2}}, \quad (5)$$

where we obtain the band diagram of the system as

$$\frac{1}{2}\mathrm{Tr}(\boldsymbol{M_U}) = \cos K\mathrm{L} = \cos(\beta L) - \frac{i\gamma_e}{-i\Delta\omega_a + \gamma_i}\sin(\beta L). \quad (6)$$

In Eq. (6), the first term represents the dispersion of the bare waveguide, while the second term describes the resonator-induced dispersive part. Due to the periodic arrangement of the resonators, the interplay between wave propagation and resonant scattering gives rise to two distinct collective dispersion mechanisms in the system. On the one hand, the localized resonances collectively hybridize into polaritonic bands, resulting in a resonance-induced polaritonic dispersion and the associated stopband centered around $\omega_a$. On the other hand, the periodic resonator scattering generates coherent Bragg backscattering under the phase-matching condition, thereby forming a Bragg-type photonic dispersion gap around the Bragg frequency $\omega_b$. If we define $\omega_0 = (\omega_a + \omega_b)/2, \delta = (\omega_b - \omega_a)/2$ as the average resonance frequency and the detuning between the two resonances, and $\Delta\omega_0 = \omega - \omega_0$ the probe detuning or the excitation detuning from average frequency $\omega_0$, then insert them into Eq. (6), we obtain

$$\cos KL = -\cos\left(\frac{\Delta\omega_0 - \delta}{v_g/L}\right) + \frac{i\gamma_e}{-i(\Delta\omega_0 + \delta) + \gamma_i}\sin\left(\frac{\Delta\omega_0 - \delta}{v_g/L}\right). \quad (7)$$

In the absence of intrinsic loss $(\gamma_i = 0)$, the resonator-induced term becomes purely real. As shown in Fig. 1(b), when there is a slight resonance detuning, two clear bandgaps related to Bragg resonance and polaritonic resonance can be observed in the band diagram, and the mode loss, reflected by value of imaginary part of the Bloch wavevector in Fig. 1(e), in polaritonic bandgap is much more pronounced compared to Bragg bandgap, indicating stronger light-matter interaction in polaritonic resonance. By tailoring the lattice period or the structural parameters of the resonators, the width of the allowed miniband confined between the Bragg gap and the resonance-induced polaritonic gap can be continuously tuned and compressed, giving rise to the EIT-like transmission effect reported in Refs. [22-24]. However, such type transparency state exists only when the two resonances remain slightly detuned ($\delta \neq 0$). As this resonance detuning approaches zero, the simultaneous vanishing of the probe detuning $\Delta\omega_0$ in the numerator and denominator maintains a nonvanishing resonator-induced scattering term, $\sim\gamma_e L/v_g$, driving the two stopbands to merge and collapse the intermediate allowed miniband. Consequently, the system becomes completely gapped at exact degeneracy, as presented in Figs. 1(c) and (f).

On the contrary, in the presence of intrinsic resonator loss $(\gamma_i \neq 0)$, the right side of the dispersion equation becomes intrinsically complex, and the imaginary part of the Bloch wavevector is no longer solely associated with the bandgap, but is also influenced by dissipation in the resonators. At the exact resonance condition, where simultaneously $\Delta\omega_0 = 0$ and $\delta = 0$, the numerator of the second term on the right-hand side of Eq. (7) vanishes while the denominator remains finite due to the nonzero $\gamma_i$, causing the resonator-induced scattering term to disappear identically. As a result, the Bloch dispersion reduces to that of the bare waveguide, i.e., $\cos(K\Lambda) = \cos(\beta L)$, as illustrated by the dispersion curves in Fig. 1(d). Consequently, the scattering channel is completely suppressed, and the low-transmission states originally associated with the Bragg and polaritonic stopbands in the intrinsic-loss-free limit evolve into a transparent state, as indicated in Fig. 1(g). The corresponding evolution of the complex Bloch wavevector over different intrinsic loss levels is further presented in Fig. S1 of [31], where the persistent dispersion crossing and the associated near-zero $\mathrm{Im}(KL)$ at the resonance frequency provide direct evidence for the formation of the transparency state. However, it is important to note that the resonators are not fully decoupled from the propagating mode. Rather, the Bragg and polaritonic resonant scatterings still persist, and it is the destructive interference between the two channels that cancel the scattering and give rise to a narrow transmission state embedded inside the broad stopband. Furthermore, the two resonant channels exhibit markedly different light–matter interactions. The polaritonic resonance is characterized by strong material absorption arising from microscopic atomic energy-level transitions, reflecting its enhanced light–matter interaction. By contrast, despite the high reflectivity at the Bragg wavelength, the Bragg resonance involves substantially weaker light–matter interaction because the electromagnetic energy is primarily stored in the propagating Bloch mode rather than dissipated through the resonators. This pronounced contrast naturally establishes a bright–dark resonance pair, closely analogous to the excited and metastable states in a Λ-type three-level EIT system, as illustrated in Fig. 2(a).

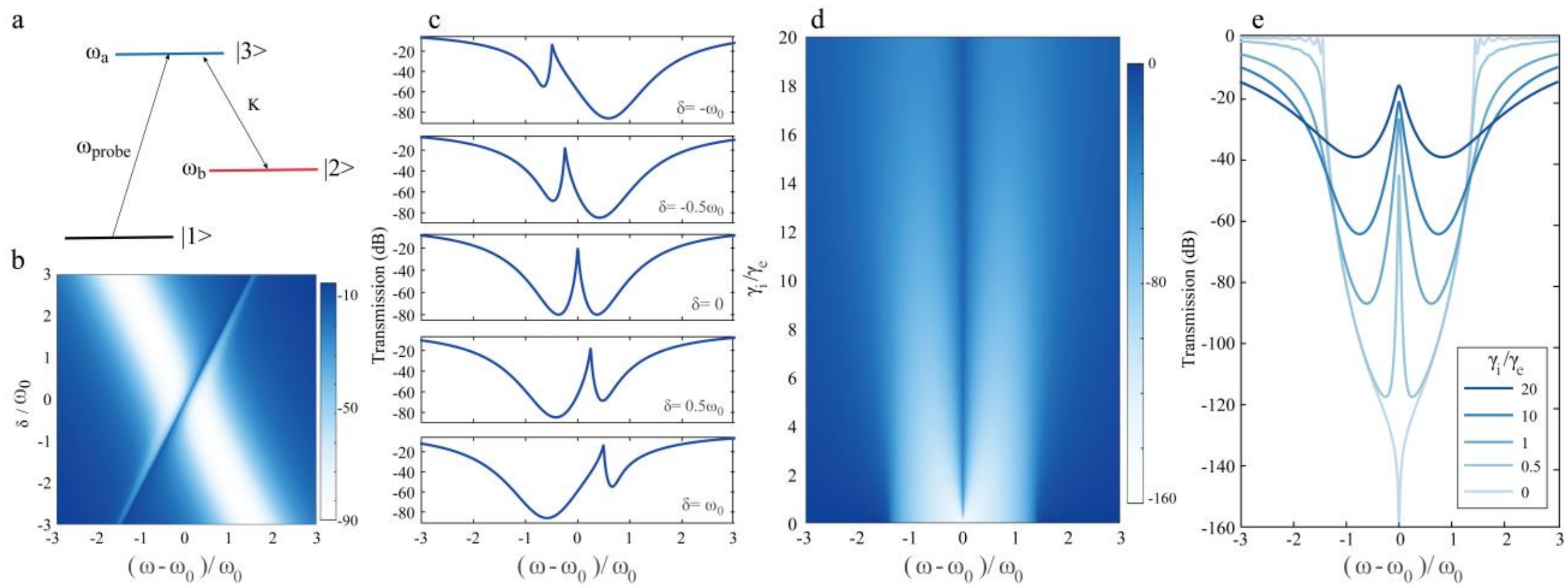

FIG. 2. (a) The mechanism analogue of bare-state picture of lambda-type three level EIT system, in which level |1> is ground state, |2> the dark state corresponds to the Bragg resonance which has very week light-matter interaction, |3> the exited state corresponds to the enhanced polaritonic resonance, which has strong light-matter interaction and can couple to dark state |2> through K. (b) The transmission map of the resonator-waveguide coupled system with N=100 periods varies with the probe frequency detuning and resonance detuning between the Bragg and polariton resonances. At exact zero detuning, a symmetric transparency window appears. Parameters of $\gamma_e = 0.1\omega_0$, $\gamma_i = 10\gamma_e$, $L\omega_0/v_g = 0.5$ are used. (c) The corresponding slices with different frequency detuning values of $\delta = [-1, -0.5, 0, 1, 0.5, 1]\omega_0$ in (b). (d) The transmission map varies with normalized frequency detuning from the averaged resonance frequency $\omega_0$ and different levels of intrinsic resonator loss $\gamma_i$ that normalized to radiative loss $\gamma_e$, at $\delta = 0$. Parameters of $\gamma_e = 0.1\omega_0$, $L\omega_0/v_g = 0.1$, N=100 are used. (e) The corresponding slices with different intrinsic resonator loss values of $\gamma_i = [0, 0.5, 1, 10, 20]\gamma_e$ in (d).

Having established the single EIT-like state in the ideal periodic structure, we now show that it remains robust against the finite-size discretization of Bloch modes and persists in structures with a finite number of resonator periods. Assuming the resonator in each unit cell has symmetric environment, that is $\Delta \boldsymbol{M_U} = 1$, we can use the theory of matrices and Chebyshev polynomials to derive the $N_{th}$ power of a unimodular matrix $\boldsymbol{M_U}$ as

$$\boldsymbol{M_N} = \boldsymbol{M_U}\frac{\sin(NK\mathrm{L})}{\sin K\mathrm{L}} - \boldsymbol{I}\frac{\sin(N-1)K\mathrm{L}}{\sin K\mathrm{L}}, \tag{8}$$

the transmission of the whole system can be readily calculated as

$$T = \left|\frac{1}{M_{N,22}}\right|^2. \tag{9}$$

Fig. 2(b) shows the transmission response of a resonator-waveguide coupled system with N=100 periods as a function of the probe-frequency detuning $\Delta\omega_0$ and the resonance detuning $\delta$ between the Bragg and polaritonic resonances, with intrinsic resonator loss $\gamma_i = 10\gamma_e$. We can observe two resonance dips in the map at large $\delta$ values: a broad dip and a narrow dip. The broad dip originates from the collective polaritonic resonance, which exhibits stronger absorption loss at the bandgap center compared to Bragg bandgap, as indicated in Fig 1(e). Its large linewidth is a direct consequence of the introduced intrinsic loss. When the two resonances are detuned slightly, their interference gives rise to characteristic asymmetric Fano line shapes in the transmission spectrum. As the resonance detuning decreases, the interference becomes increasingly balanced, eventually producing an ultra-narrow and nearly symmetric EIT window at zero detuning. The evolution from Fano resonance to EIT is more clearly observed in the transmission spectra extracted from Fig. 2(b) at several representative detuning values, as presented in Fig. 2(c).

To reveal how intrinsic resonator loss gives rise to the EIT-like state, Fig. 2(d) presents the transmission map at zero resonance detuning ($\delta = 0$) as functions of the probe-frequency detuning $\Delta\omega_0$and the intrinsic spectra resonator loss $\gamma_i$, while representative transmission are shown in Fig. 2(e). In the absence of intrinsic loss ($\gamma_i = 0$), a well-defined forbidden band is observed, featuring an ultranarrow transmission dip at its center. This dip originates from the collective polaritonic resonance and represents the strongest absorption state within the forbidden band, in sharp contrast to conventional distributed-feedback cavities, where the band-center transmission minimum is dominated by Bragg reflection rather than absorption. Once intrinsic resonator loss is introduced, the strongly absorbing gap-center mode evolves into a narrow transparency mode. At low loss levels, the forbidden and allowed bands remain well defined, while a narrow transparency window already emerges around zero probe detuning. As the intrinsic loss further increases, the transmission within the spectral region originally corresponding to the forbidden band is progressively enhanced, indicating a continuous reduction of the forbidden-band attenuation, consistent with the decrease of $\mathrm{Im}(KL)$ shown in Fig. S1(b) in [31]. In contrast, the transmission in the previously allowed band gradually decreases because of the increasing dissipative background introduced by the resonator loss. Consequently, the transmission contrast between the original forbidden and allowed bands is continuously reduced, causing the band edges to become progressively blurred. At sufficiently high intrinsic loss, the distinction between the forbidden and allowed bands disappears, leaving a broad lossy background with an ultranarrow transparency peak centered at $\Delta\omega_0 = 0$. This increase in transparency, however, comes at the expense of a broader linewidth, indicating a trade-off between peak transmission and spectral selectivity.

Encouragingly, this broadening can be offset by employing a larger number of resonator periods, thereby enabling narrower transparency features. Figure 3(a) presents the EIT-like features in the transmission spectra of the resonator-waveguide coupled system for period numbers ranging from N=100 to 500 at zero resonance detuning. As the number of periods increases, the contrast between the transparency peak and the surrounding transmission valleys becomes increasingly pronounced, while the linewidth of the EIT window is significantly narrowed. To quantitatively characterize this behavior, Fig. 3(b) shows the quality (Q) factor of the EIT mode as a function of the period number for several intrinsic loss levels. For a fixed intrinsic loss, the Q factor generally increases with the number of periods, exhibiting an approximately linear dependence in the large N regime. In contrast, for a fixed number of periods, lower intrinsic loss leads to a higher Q factor, indicating that both the structural length and the resonator dissipation jointly determine the spectral selectivity of the transparency mode. Notably, a Q factor approaching $10^6$ can be achieved for N=1000 and $\gamma_i = 0.1\gamma_e$, demonstrating exceptional mode selectivity enabled by distributed resonant feedback, thereby enabling ultranarrow-linewidth single-frequency laser operation.

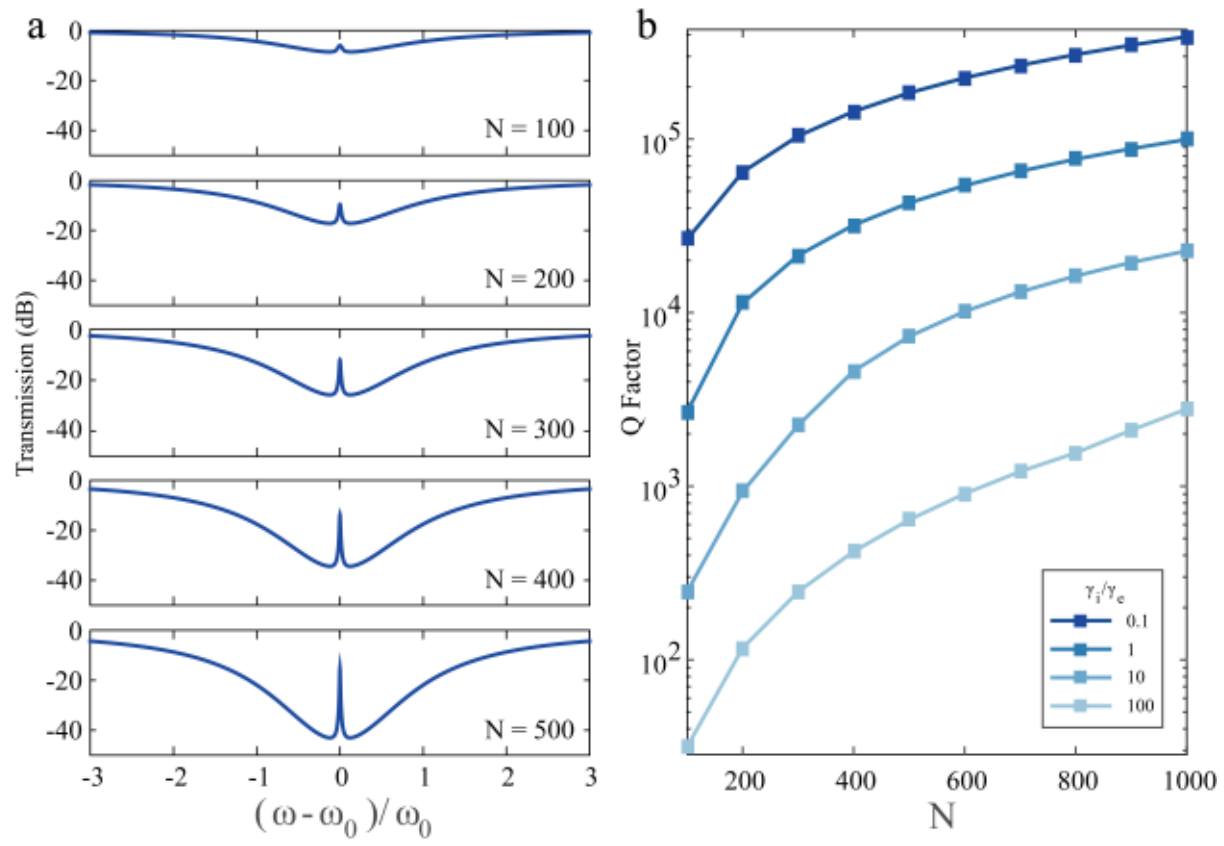

FIG. 3. (a) EIT-like features in the transmission spectra of the resonator-waveguide coupled system for different numbers of periods ranging from N=100 to 500 with $\gamma_i = 10\gamma_e$. (b) Corresponding Q factors of the EIT windows as a function of the number of periods (N=100 to 1000) for several intrinsic resonator loss levels, $\gamma_i = [0.1, 1, 10, 100]\gamma_e$. Common parameters $\delta = 0$, $\gamma_e = 0.01\omega_0$ and $L\omega_0/v_g = 0.5$ are used in the calculations.

To demonstrate its practical implementation, we design a 1-mm-long laser cavity operating at 1550 nm by integrating a plasmonic resonator array with an InP/InGaAsP multiple-quantum-well waveguide through side coupling. Each resonator supports only a single localized plasmonic resonance ($Q \sim 10$). With an effective refractive index of approximately 3.229, the corresponding lattice period is 240 nm, yielding approximately 4166 periods. To reveal the intrinsic modal characteristics enabled by distributed resonant feedback, facet reflections are neglected throughout the analysis. Fig. 4(a) presents the transmission map of the cavity over an ultra-broad spectral range of 600 nm as the material gain varies from –40 to 200 cm$^{-1}$. A series of lasing poles are observed, corresponding to the quantized Bloch resonances supported by the finite periodic structure. Around the target wavelength of 1550 nm, however, the proposed loss-enabled EIT mechanism gives rise to an isolated lasing pole with a substantially lower threshold than all neighboring modes. As shown in Fig. 4(b), this pole reaches the lasing condition at a modal gain of only 16.6 cm$^{-1}$. The enlarged view in Fig. 4(c) further reveals a pronounced gain separation between the fundamental lasing pole and the nearest competing poles, indicating exceptionally strong mode discrimination. Such a large threshold margin effectively suppresses mode competition and ensures robust single-mode operation even in the presence of broadband optical gain. Fig. 4(d) shows the corresponding lasing spectrum at threshold. The lasing peak coincides with the ultranarrow transparency resonance of the passive cavity, confirming that the lasing action is directly supported by the distributed resonant feedback established through the loss-enabled EIT state rather than conventional facet reflections. It is worth noting that the above analysis neglects facet reflections to isolate the intrinsic distributed resonant feedback mechanism. When realistic facet reflections are included, the lowest-threshold lasing mode remains associated with the EIT resonance and exhibits a threshold even lower than that of a conventional Fabry–Pérot cavity with the same cavity length, as indicated in Fig.S2 of [31].

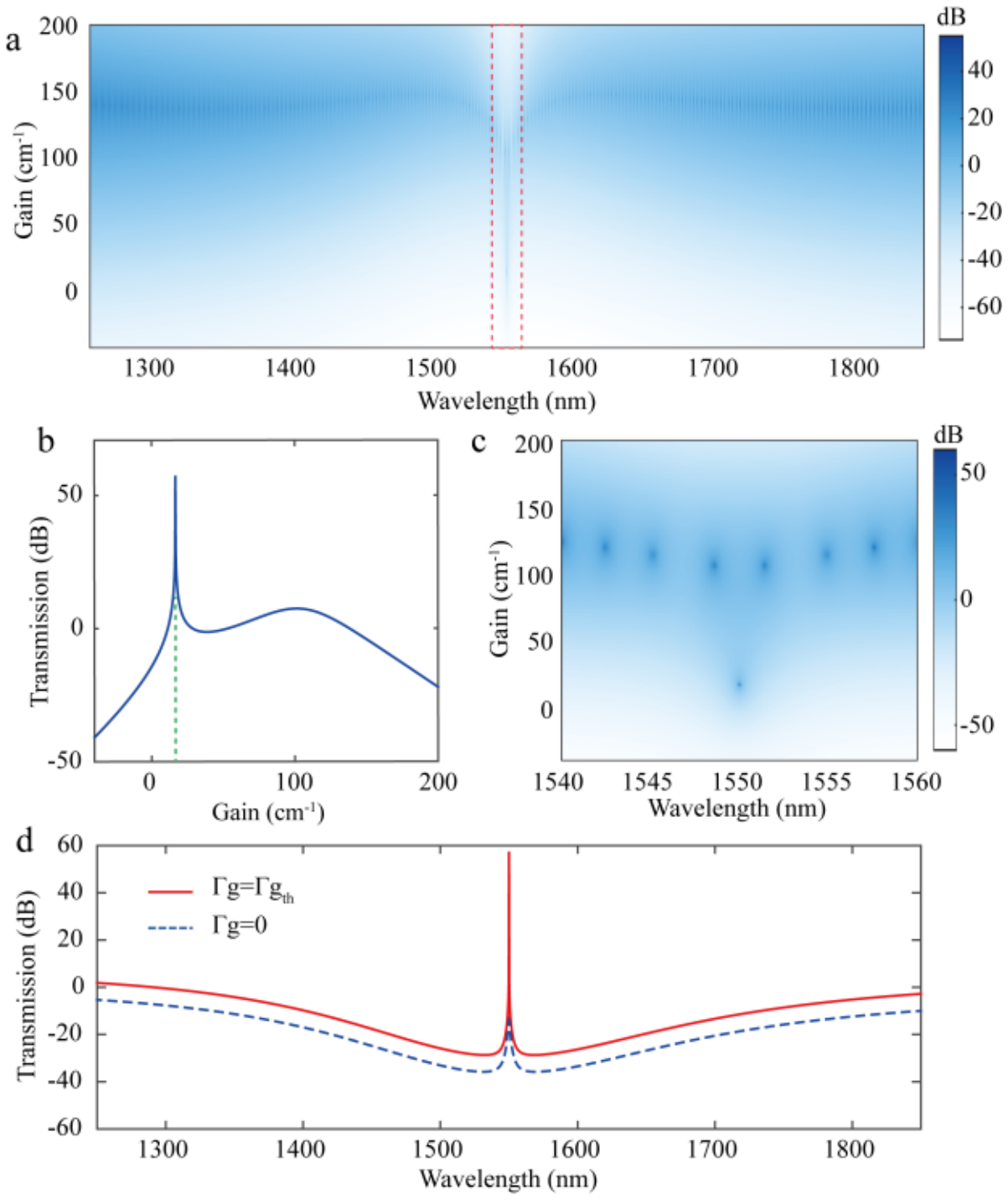

FIG.4. (a) Transmission map of a periodic waveguide-resonator coupled cavity with a unit cell consisting of a plasmonic resonator and an InP/InGaAsP waveguide. Designed for operation at 1550 nm, the map is calculated over an ultra-broad spectral range of 600 nm under varying material gain levels from –40 to 200 cm⁻¹, with facet reflections excluded. (b) Transmission slice at the target wavelength of 1550 nm, showing the emergence of a lasing pole (indicated by the green dashed line) at a modal gain of 16.6 cm⁻¹. (c) Enlarged view of the region enclosed by the red dashed box in (a). (d) Lasing spectrum corresponding to the lowest-threshold lasing pole. The blue dashed curve denotes the passive transmission spectrum of the cavity. The parameters are $\gamma_e = 0.0001\omega_0$, $\gamma_i = 0.1\omega_0$, $L\omega_0/v_g = 0.5$, and $N = 4166$.

By revealing the physical origin of loss-enabled EIT in periodic resonator arrays, this work establishes a new paradigm for exploiting dissipation as a constructive resource in collective photonic systems. More importantly, it extends the conventional role of EIT from passive spectral transparency to active single-frequency laser operation, providing a fundamentally new framework for engineering coherent light generation and integrated non-Hermitian photonic functionalities.

This work was supported by Guangdong Provincial Quantum Science Strategic Initiative (GDZX2502007), Guangdong Basic and Applied Basic Research Foundation (2023A1515012793, 2024B1515020117) and the start-up funding from the Quantum Science Center of the Guangdong-Hong Kong-Macao Greater Bay Area.

* liangyaoyao@quantumsc.cn ;
† wangfeihu@quantumsc.cn